\documentclass[pra,twocolumn,superscriptaddress]{revtex4-1}
\usepackage{amsmath, amssymb,graphicx,color,bm,epstopdf}
\usepackage[dvipsnames]{xcolor}
\usepackage{bbold}
\usepackage{braket}
\usepackage{comment}
\usepackage[unicode=true,colorlinks=true,citecolor=blue,urlcolor=blue]{hyperref}
\usepackage{graphicx}

\newcommand{\beml}{\begin{subequations}}
\newcommand{\eml}{\end{subequations}}
\newcommand{\beq}{\begin{eqnarray}}
\newcommand{\eq}{\end{eqnarray}}
\newcommand{\ba}{\begin{array}}
\newcommand{\ea}{\end{array}}
\newcommand{\bpm}{\begin{pmatrix}}
\newcommand{\epm}{\end{pmatrix}}
\newcommand{\bc}{\begin{cases}}
\newcommand{\ec}{\end{cases}}

\renewcommand {\phi}{{\varphi}}
\newcommand {\rmi}{{\rm i}}

\newcommand {\rmd}{{\rm d}}

\newcommand {\e}{{\rm e}}

\usepackage[utf8]{inputenc}
\begin{document}
\title{Localized vibrational modes in waveguide quantum optomechanics with spontaneously broken $\mathcal{PT}$ symmetry }
\author{Alexander Poshakinskiy}
\affiliation{Ioffe Institute, 194021, St. Petersburg, Russia}

\author{Ivan Iorsh}
\affiliation{Department of Physics and Technology, ITMO University, St. Petersburg, 197101, Russia}
\author{Alexander Poddubny}
\affiliation{Ioffe Institute, 194021, St. Petersburg, Russia}
\affiliation{Department of Physics and Technology, ITMO University, St. Petersburg, 197101, Russia}
\begin{abstract}
We study theoretically  two vibrating quantum emitters trapped near a one-dimensional waveguide and interacting with propagating photons. We demonstrate, that  in the regime of strong optomechanical interaction the light-induced coupling of emitter vibrations can lead to formation of spatially localized vibration modes, exhibiting parity-time ($\mathcal{PT}$) symmetry breaking. These localized vibrations can be interpreted as  topological defects in the quasiclassical energy spectrum.
\end{abstract}

\maketitle
\section{Introduction}\label{sec:intro} 
Interaction of localized quantum emitters with  photons, propagating in the waveguide,  is now actively studied, both theoretically and experimentally~\cite{roy2017,KimbleRMP2018,sheremet2021waveguide}. On the applied side, this new platform of waveguide quantum electrodynamics offers novel opportunities to generate and manipulate quantum light. On the fundamental side, the long-ranged  waveguide-induced coupling between the emitters allows one to explore novel many-body quantum phases, mediated by quantum interactions. Novel degrees of freedom are opened in the regime of {\it waveguide quantum optomechanics}, when the emitters can move in space. This platform can be potentially  realized with optically trapped atoms and experimental demonstrations of mechanically mediated microwave frequency conversion~\cite{Teufel2016} and optomechanical nonreciprocity with superconducting qubits~\cite{Peterson2017} are already available. The Hilbert space of the system is spanned by three types of excitations: (i) propagating photons, (ii) emitter excitons, and (iii) vibration quanta~\cite{Chang2013,Lesanovsky2020,dong2021unconventional}. Their coupling and interactions lead to very  interesting effects even in the absence of mechanical motion quantization~\cite{Burillo2020}, for example, Ref.~\cite{Chang2013} has predicted optically-induced
self-organization of the atom coordinates along a waveguide. 

\begin{figure}[!b]
    \centering
    \includegraphics[width=0.9\columnwidth]{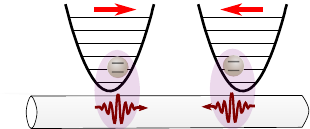}
    \caption{Schematics of the setup: two vibrating atoms in the waveguide coupled  by propagating photons.}
    \label{fig:1}
\end{figure}

In our  recent theoretical work \cite{Iorsh2020}, we have extended the above studies to the quantum domain by considering the quantized mechnicanical vibrations of two atoms coupled by single polaritonic excitation, see Fig.~\ref{fig:1}. We have predicted that this setup features ultrastrong optomechanical coupling~\cite{Casanova2018,FriskKockum2019} and can manifest parity-time symmetry breaking. The symmetry breaking  can be potentially observed by studying the inelastic scattering of waveguide photons  with phonon emission or absorption.  In the following work \cite{Iorsh2020b}, these results have been extended to three atoms interacting with a chiral waveguide, where photons can propagate only in one direction, and quantum phase transitions with $\mathbb Z_3$ symmetry breaking have been predicted.  Importantly, the study of Refs.~\cite{Iorsh2020} and~\cite{Iorsh2020b}  was focused only on the regime with just several vibration quanta. To the best of our knowledge, the waveguide quantum optomechanics in the regime when the number of vibration quanta is moderate,  corresponding to transition between the ultra-quantum regime of Ref.~\cite{Iorsh2020} and  semiclassical regime of Ref.~\cite{Chang2013},  remains uncharted. 

Here,  we consider a pair of vibrating atoms coupled to the waveguide in the regime of many vibration quanta. In this intermediate regime, we reveal the formation of a new type of vibration states which are strongly localized in space due to the optomechanical interaction and can exhibit parity-time ($\mathcal{PT}$) symmetry breaking~\cite{Feng2017}. Our numerical calculations are corroborated by a quasiclassical theoretical analysis of the vibration energy quantization.

The rest of the work is organized as follows. Section~\ref{sec:model} sums up the model and the calculation approach. 
Importantly, the considered problem is strongly non-Hermitian due to the possibility of the radiative losses in the waveguide.  We first ignore these losses and analyze in Sec.~\ref{sec:hermitian} the Hermitian regime that already allows us to explain the essence of the quasiclassical localization. Next, in Sec.~\ref{sec:non-hermitian} we take into account the radiative losses and demonstrate how the localized modes exhibit $\mathcal{PT}$ symmetry breaking. The results are summarized in Sec.~\ref{sec:summary}.

\section{Model}\label{sec:model} 
The system under consideration is shown in Fig.~\ref{fig:1} and is described by the following Hamiltonian
\begin{align}\label{eq:H1}
\hat{H}=\sum_k \omega_k c_k^{\dagger}c_k +\sum\limits_{j=1}^2 \omega_x\sigma_j^{+}\sigma_j+\sum\limits_{j=1}^2\Omega a_j^{\dagger}a_j +\hat{H}_{\rm int}\:.
\end{align}
Here,  $c_k$ and $a_j$ are the annihilation operators for photons and atom vibrations, $\sigma_j$ is the lowering operator for the two-level atom, $\omega_k=v|k|$ is the frequency of the waveguide mode, $k$ is the photon wave vector,  $v$ is the mode velocity, $\omega_x$ is the atom resonance frequency, and $\Omega$ is the frequency of atom mechanical motion in the optical trap. The interaction Hamiltonian is given by
\begin{align}\label{eq:H2}
\hat{H}_{\rm int}=g\sum_{k}\sum\limits_{j=1}^2\left[\sigma_j^{\dagger}c_k^{\vphantom{\dagger}}\e^{ik[z_j+u_0(a_j+a_j^{\dagger})]}+\mathrm{H.c.}\right],
\end{align}
where  $z_j$ are the equilibrium atom coordinates, $u_0$ is the vacuum-displacement amplitude, and we assumed rotating wave and dipole approximations for the light-atom interaction.   Even after these approximations, the problem is still quite challenging. Namely, the Hamiltonian Eq.~\eqref{eq:H1},\eqref{eq:H2} is strongly nonlinear in the vibration operators $a_j$ and it is unclear how to solve it in the general many-body case. Importantly, interaction~\eqref{eq:H2} conserves the total number of photons and atom excitations (but not the number of vibrations).  Here, we restrict ourselves to a subspace  of single photon or atom excitation, when the problem can be considerably simplified, as discussed in detail in our previous work \cite{Iorsh2020}. For the sake of completeness, we recast below the procedure that mostly follows the Supplemental Materials of Ref.~\cite{Iorsh2020}.

Our first step is to integrate out the photonic degrees of freedom $\hat{c}_k$. This can be done by a path integral approach~\cite{Shi2011} and the result reads
\begin{align}
&\hat{H}_{\rm eff}=\sum_{j=1}^2 \omega_x\sigma_j^{+}\sigma_j+\sum_{j=1}^2\Omega a_j^{\dagger}a_j \label{eq:H_eff}\\&-\rmi\Gamma_0\sum_{j,l=1}^2\sigma_j^{+}\sigma_le^{\rmi q|z_j-z_l|}\e^{\rmi qu_0 \,\text{sign}(z_j-z_l)\,(a_j+a_j^{\dagger}-a_l-a_l^{\dagger})},  \nonumber
\end{align}
where $\Gamma_0=g^2/v$ is the radiative decay rate of a single qubit and we used the Markovian approximation neglecting the frequency dispersion in the phase factor, $k\approx q \equiv \omega_x/v$. An alternative way to obtain the same Hamiltonian~\eqref{eq:H_eff} is to start from the conventional waveguide-QED  Hamiltonian~\cite{Molmer2019,Ke2019} 
\begin{align}\label{eq:WQED}
&\sum_{j=1}^2 \omega_x\sigma_j^{+}\sigma_j+\sum_{j=1}^2\Omega a_j^{\dagger}a_j -\rmi\Gamma_0\sum_{j,l=1}^2\sigma_j^{+}\sigma_le^{\rmi q|z_j-z_l|}
\end{align}
and replace the coordinates of static atoms $z_j$ with their corresponding dynamical values $z_j+ u_0(a_j+a_j^{\dagger})$, where we assume that  $u_0 \lesssim |z_j-z_l|$. 

Next,  we project the Hamiltonian Eq.~\eqref{eq:H_eff} onto the subspace with a single atom  excitation. This is possible because  the operator of the total number of atom excitations $\mathcal{N}=\sum_{j}\sigma_{z,j}$ commutes with the Hamiltonian $\hat{H}_{\rm eff}$. The resulting effective Hamiltonian for atomic excitations coupled to the vibrations can be represented as a $2\times 2$ matrix, with each matrix element depending only on phononic degrees of freedom.
We also exploit the mirror symmetry of the problem and 
 introduce a new rotated basis comprising a symmetric and antisymmetric superposition of the atom excitations and vibrations : \[
 |\pm\rangle =\frac{1}{\sqrt{2}}(|L\rangle\pm |R\rangle), \quad 
 \hat{x}_{s(d)}=\frac{\hat{x}_2\pm \hat{x}_1}{\sqrt{2}}\:,
 \]
 where $|L\rangle = \sigma_1^\dag|0\rangle$ and $|R\rangle= \sigma_2^\dag|0\rangle$ are the states with left or right atom excited, 
 $\hat{x}_i=(a_i+a_i^{\dagger})/\sqrt{2}$ are the atom displacements  measured in units of $u_0$. In the new basis $|\pm\rangle$, the matrix of the  effective Hamiltonian assumes the diagonal form, with the diagonal elements given by
\begin{align}
\hat{H}_{\rm eff}^{\pm}=\Omega a_s^{\dagger}a_s^{\vphantom{\dagger}} +\Omega a_d^{\dagger}a_d^{\vphantom{\dagger}} -\rmi\Gamma_0\left(1 \pm e^{\rmi\phi} e^{\rmi\eta \hat{x}_d}\right),\label{eq:H_d0} 
\end{align}
where $\phi=q|z_1-z_2|$ and $\eta=2qu_0$.  The symmetric vibrations $x_s$ do not affect the distance between the qubits and are decoupled from the atomic excitations so we will disregard them in what follows. 
In this work, we focus on the part of the Hamiltonian corresponding to the symmetric atom excitation $|+\rangle$.  {Consideration of the $|-\rangle$ state gives similar results.}
Following Ref.~\cite{Iorsh2020}, we find the spectrum of Eq.~\eqref{eq:H_d0}  by using the real-space representation of the mechanical motion. We  introduce the  coordinate variable $x_d$ (measured in units of $\sqrt{2}u_0$) and  make the transformation $\hat{x}_d\rightarrow x_d,\hat{\dot{x}}_d\rightarrow-\rmi\Omega\, (\rmd/\rmd x_d)$. The resulting Hamiltonian then reads~\cite{Iorsh2020}:
\begin{align}~\label{eq:H_d}
&\hat H_{\rm eff}^{+} =
\frac{\Omega}{2}\left(-\frac{\rmd^2}{\rmd x_d^2}-1\right)-\rmi\Gamma_0+V(x_d)\:,\\
&
V(x_d)=\frac{\Omega x_d^2}{2}+\rmi\Gamma_0\e^{\rmi\phi}\e^{\rmi\eta x_d} \nonumber
\end{align}
and can be straightforwardly diagonalized using the finite difference discretization scheme. Similarly to the WQED Hamiltonian  Eq.~\eqref{eq:WQED},  Eq.~\eqref{eq:H_d}  is non-Hermitian due to the radiative losses into the waveguide. However, in what follows we will focus on the case when the relative distance between the atoms is a quarter of wavelength, i.e., $\varphi=\pi/2$, so that $-\rmi\Gamma_0\e^{\rmi\phi}\equiv \Gamma_0$ and the dissipative coupling for stationary atoms is suppressed.

\section{Hermitian regime}\label{sec:hermitian} 
\begin{figure}[!t]
    \centering
    \includegraphics[width=0.99\columnwidth]{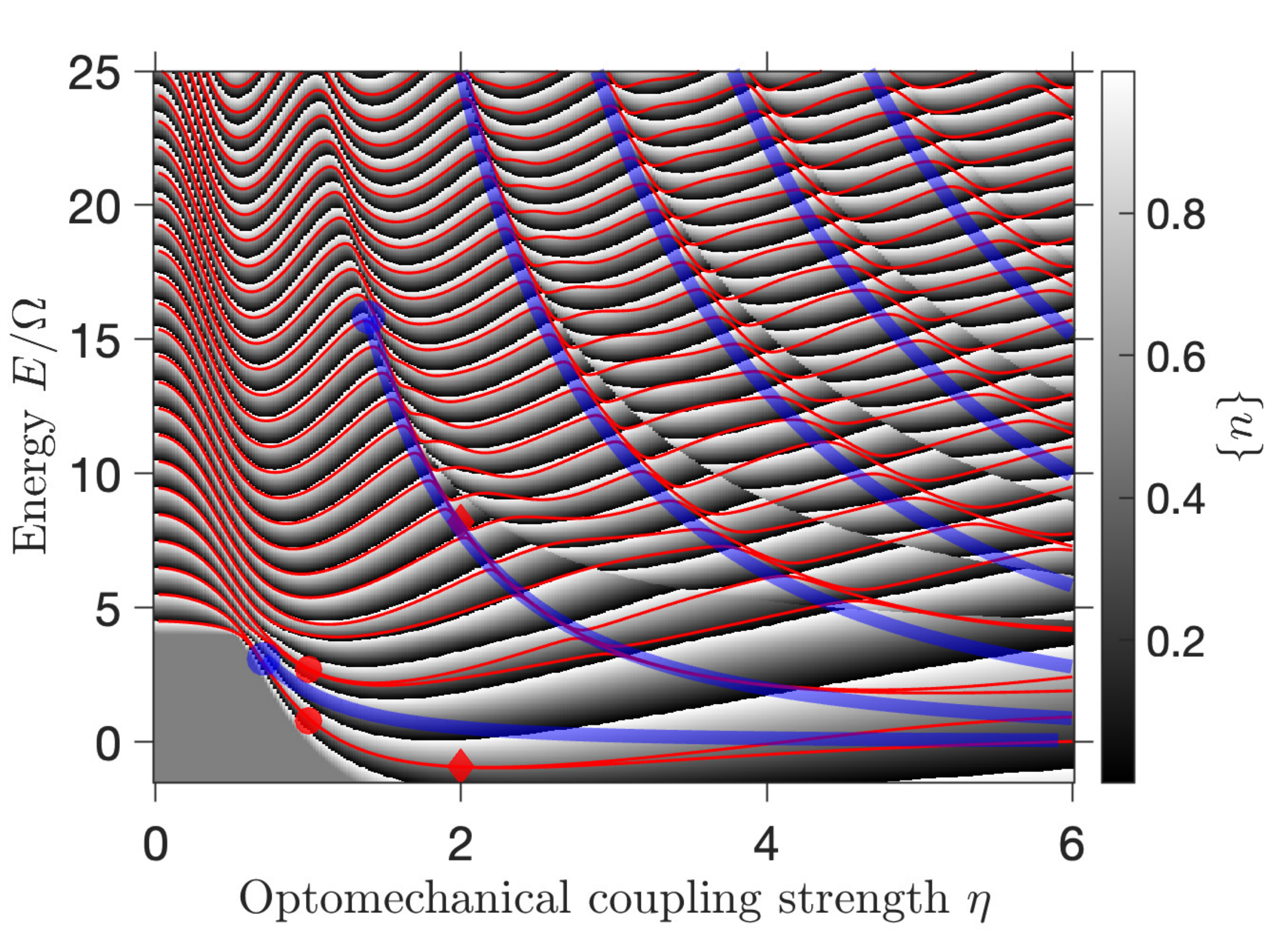}
    \caption{Dependence of the spectrum on the optomechanical coupling strength calculated neglecting the dissipative terms in the Hamiltonian Eq.~\eqref{eq:H_d}. Calculation has been performed for $\Gamma_0/\Omega=4$. Grayscale shading shows the fractional part of the quasiclassical phase $\{n\}$ calculated after Eq.~\eqref{eq:n}. Red circles and diamonds show the energies of the eigenstates that are illustrated in Fig.~\ref{fig:3}.  
    Thick blue lines are calculated following  analytical Eqs.~\eqref{eq:eta0}--\eqref{eq:bohr1}.     }
    \label{fig:2}
\end{figure}
In this section we neglect the anti-Hermitian part  of the Hamiltonian Eq.~\eqref{eq:H_d}, which describes the dissipative coupling, i.e., replace $V(x_d)$ in Eq.~\eqref{eq:H_d} with \[V'(x_d) \equiv \text{Re\,} V(x_d) =\Omega x_d^2/2+ \Gamma_0\cos (\eta x_d)\:.\] Shown in Fig.~\ref{fig:2} by red lines is the energy spectrum of such Hamiltonian as a function of the dimensionless parameter $\eta$ describing the optomechanical coupling strength. When $\eta=0$, the optomechanical coupling leads just to an overall shift of the spectrum and the energy levels  form an equidistant ladder $E_n=n\Omega+\Gamma_0$, $n=0,1,2\ldots$, characteristic for a harmonic oscillator. Increase of the coupling strength transforms the spectrum and leads to mixing and avoided crossing between different oscillator states. In the  limit of large coupling, $\eta\gg 1$, the fast oscillating term $\propto \cos \eta x_d$ in the potential $V'(x_d)$  can be neglected and the energy spectrum again tends to an equidistant ladder.

The calculation demonstrates that in the intermediate region of $\eta$, the optomechanical coupling potential
$\propto \cos \eta x_d$ introduces a periodical modulation into the energy levels. The amplitude of the energy modulation is determined {by the ratio of the modulation period $2\pi/\eta$ and the characteristic size of the oscillator wavefunction  $\propto \sqrt{E}$}. Another important result seen from Fig.~\ref{fig:2} is the appearance of additional pairs of almost degenerate  energy levels when the optomechanical coupling  exceeds a certain threshold. For example, after the threshold {$\eta_1\approx 1.4$} the pair of levels with {$E_1\approx 16\Omega$} appears in the energy spectrum (see the blue circle in Fig.~\ref{fig:2}). After the threshold is crossed, the new pair of levels stays in the energy spectrum and its energy decreases {$\propto 1/\eta^2$}.

\begin{figure}[!tb]
    \centering
    \includegraphics[width=0.97\columnwidth]{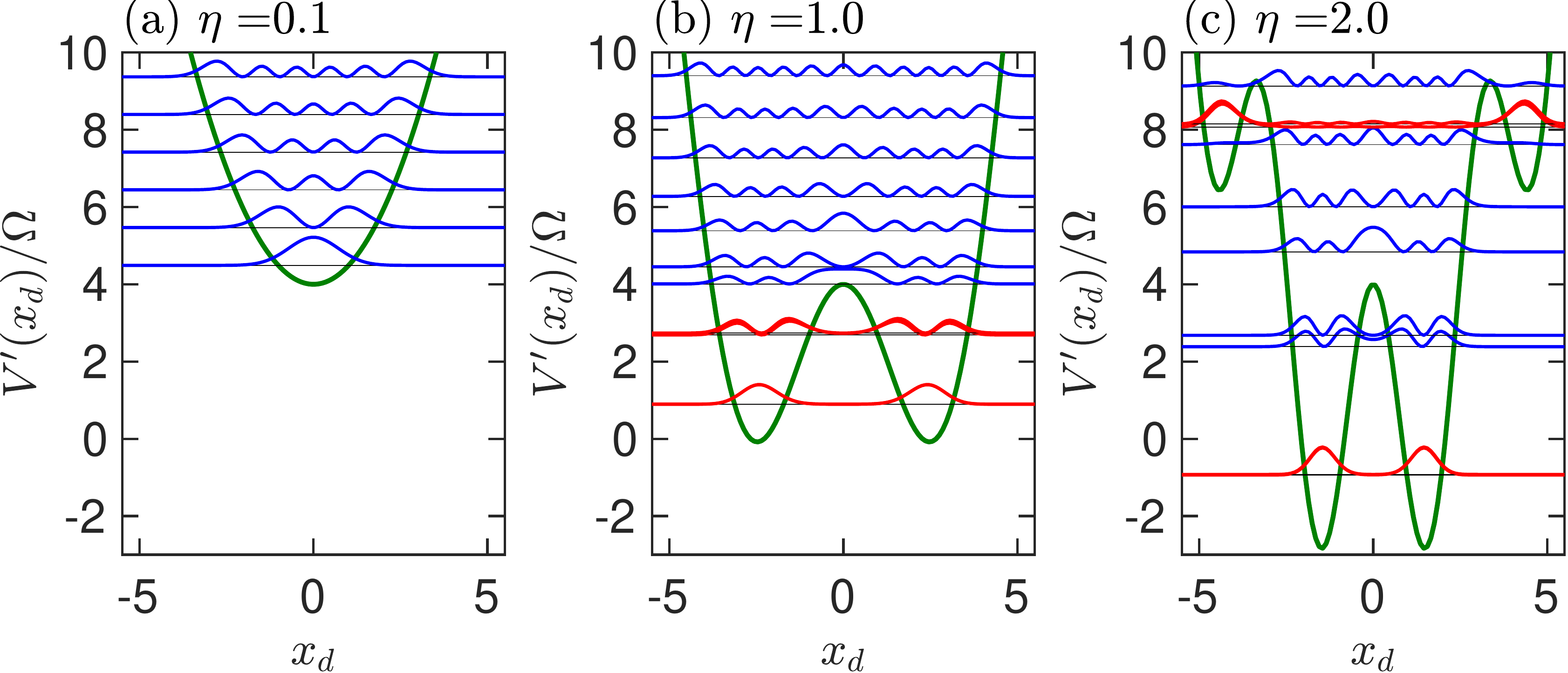}
    \caption{Eigenfunctions $|\psi^2_n(x_d)|$ of the Hamiltonian Eq.~\eqref{eq:H_d}  calculated neglecting the dissipative terms (black and red curves), the potential $V'(x_d)$ (green curve) and the energy levels (thin black horizontal lines), corresponding to three specific values of optomechanical coupling strength $\eta$. Red color shows the localized states marked by red circles and diamonds in Fig.~\ref{fig:2}.
  }
    \label{fig:3}
\end{figure}

Shown in Fig.~\ref{fig:3} are the calculated eigenfunctions for different values of optomechanical coupling strength $\eta$. Most of the eigenfunctions are delocalized in $x_d$, similarly to the oscillator wavefunctions at $\eta=0$. Interestingly,
the eigenfunctions corresponding to the discussed above pairs of almost degenerate states (marked by red color) appear to be localized  in vicinity of certain $x_d$ and $-x_d$. 
 In particular, the calculation for $\eta=2$ shown in Fig.~\ref{fig:3}c reveals a pair of almost degenerate states with {$E/\Omega\approx -1 $ localized close to $x_d =\pm 1.5$ and another pair of states with $E/\Omega\approx 8 $ localized close to $x_d = \pm 4.4$}. 
 
 The qualitative origin of the extra levels in the energy spectrum is evident from the shape of the potential $V'(x_d)$, plotted in Fig.~\ref{fig:3} by green curve. At small coupling $\eta$, the potential is close to parabolic and has only one minimum at $x_d=0$.  When  $\eta$ is increased, the oscillating term $\Gamma_0\cos\eta x_d$ leads to appearance of additional potential minima in the potential. This minima act as potential wells (with one barrier being semitransparent) where the new eigenstates can be localized.

More insight in the energy spectrum can be obtained in the quasiclassical approximation. 
For any energy $E$ we introduce the classical turning points $ \tilde x_d$, satisfying the equation $V'(\tilde x_d) = E$.
Then, we calculate the quasiclassical phase
\begin{equation} \label{eq:n}
n= \frac{1}{\pi}\int\limits_{-\tilde x_d}^{\tilde x_d} \sqrt{2[E-V'(x_d)]}\,  \rmd x_d \, {-\frac12}\:.
\end{equation}
According to Bohr-Sommerfeld quantization rule, quantized energy levels should correspond to integer $n$.  
The grayscale shading in Fig.~\ref{fig:2} shows the fractional part of the  quasiclassical phase $\{ n \}$ calculated numerically and  plotted depending on the optomechanical coupling strength $\eta$ and on the energy $E$. The calculation demonstrates that the phase discontinuities well correspond to the  exact  energy levels of the potential, especially in the applicability region of the quasiclassical approximation, $E\gg \Omega$. Moreover, this quasiclassical approach also captures the uncovered localized states. Appearance of additional potential minima leads to the presence of more than two turning points. The ambiguity in the choice of $ \tilde x_d$ results in the branching of the $n(\eta,E)$ function. We resolve the ambiguity by choosing the  turning points with the smallest absolute value, which results in appearance of branch cuts. {The branch cut position corresponds well to the energy of localized states. The point  of the localized state appearance is manifested in the phase pattern by a feature that resembles  fork defects for the interference fringes}. 

In order to find the approximate energy of the localized states analytically, we 
write the condition for the appearance of the new potential minima as 
\begin{align}
&\frac{\rmd V'(x_d)}{\rmd x_d} = \Omega x_d-\Gamma_0\eta\sin\eta x_d =0 \,, \label{eq:dq}\\
&\frac{\rmd^2V'(x_d)}{\rmd x_d^2} = \Omega- \Gamma_0\eta^2\cos \eta x_d = 0\,. \label{eq:dq2}
\end{align} 
Combining  these two equations we find that $\eta x_d = \pm A_n$ where $A_n$ is the set of non-negative solutions of the equation $
\tan A_n=A_n $ satisfying the additional condition $\cos A_n > 0$. The first of them is $A_0 =0$ and the other solutions with the accuracy better that $10^{-3}$ can be approximated as 
\begin{align}
A_n \approx  2\pi n +\pi/2 - \frac{1}{2\pi n + \pi/2}  \quad (n \geq 1).
\end{align}
Then, using Eq.~\eqref{eq:dq2} we find that the threshold values $\eta_n$ for the localized states appearance read
\begin{align}\label{eq:eta0}
&\eta_0 = \sqrt{\frac{\Omega}{\Gamma_0}} \,,\\
&\eta_n = \sqrt{\frac{\Omega}{\Gamma_0 \cos A_n}} \approx \sqrt{\frac{ 2\pi \Omega(n+1/4)}{\Gamma_0}} \quad (n \geq 1) \nonumber
\end{align}
and the corresponding energies are
\begin{align}
& E_0 = \Gamma_0  \,,\\
& E_n = \Gamma_0  \cos A_n (1 + A_n^2/2) \approx \pi \Gamma(n+1/4) \quad (n \geq 1). \nonumber
\end{align}
At $\eta > \eta_n$, the potential $V'(x_d)$ has $n+1$ pairs of additional minima. Depending on their depth and width, the number of states localized inside them can vary, e.g., in Fig.~\ref{fig:3}(c) both the first and second pair of additional minima host one pair of localized states (marked red), whereas  in Fig.~\ref{fig:3}(b) the first pair of additional minima hosts two pairs of states. The energies of the localized states are bounded by the corresponding potential minimum and the lowest adjacent  potential maximum. {A good estimation for the average energy of the states localized in a certain minimum is provided by the potential in inflection point between the minimum and maximum of this potential,  which can be found using Eq.~\eqref{eq:dq2} and reads
\begin{equation}\label{eq:bohr1}
\bar E_n(\eta) \approx \frac{\Omega}{2\eta^2}\left[\left(2\pi n+\arccos\frac{\Omega}{\Gamma_0\eta^2}\right)^2+2\right] \quad (\eta \geq\eta_n).
\end{equation}}
Thick blue curves in Fig.~\ref{fig:2} are plotted after Eq.~\eqref{eq:bohr1} and demonstrate reasonable agreement with the energies of numerically calculated edge states.

\section{Parity-time symmetry breaking}\label{sec:non-hermitian} 
\begin{figure}[t]
    \centering
    \includegraphics[width=\columnwidth]{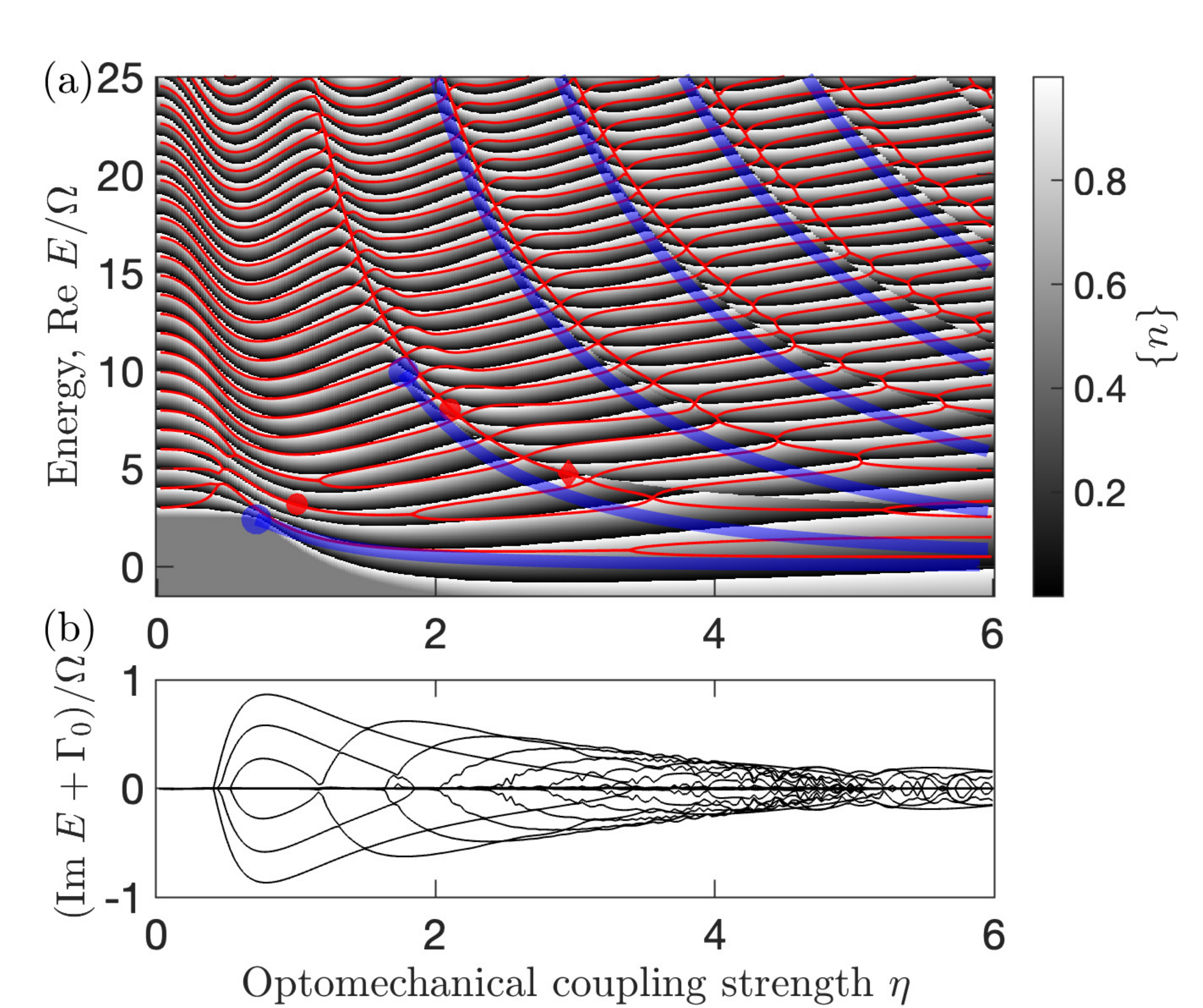}
    \caption{
    Dependence of the real part of the spectrum on the optomechanical coupling strength calculated  including the dissipative terms in the Hamiltonian Eq.~\eqref{eq:H_d}. Calculation has been performed for $\Gamma_0/\Omega=4$. Grayscale shading shows the fractional part of the quasiclassical phase $\{n\}$. Red circles and diamonds show the energies of the eigenstates that are illustrated in Fig.~\ref{fig:5}.      Thick blue lines are calculated following  analytical Eqs.~\eqref{eq:eta0}--\eqref{eq:bohr1}.    (b) Imaginary part of the energy spectrum.
    }
    \label{fig:4}
\end{figure}
\begin{figure}[!b]
    \centering
    \includegraphics[width=0.97\columnwidth]{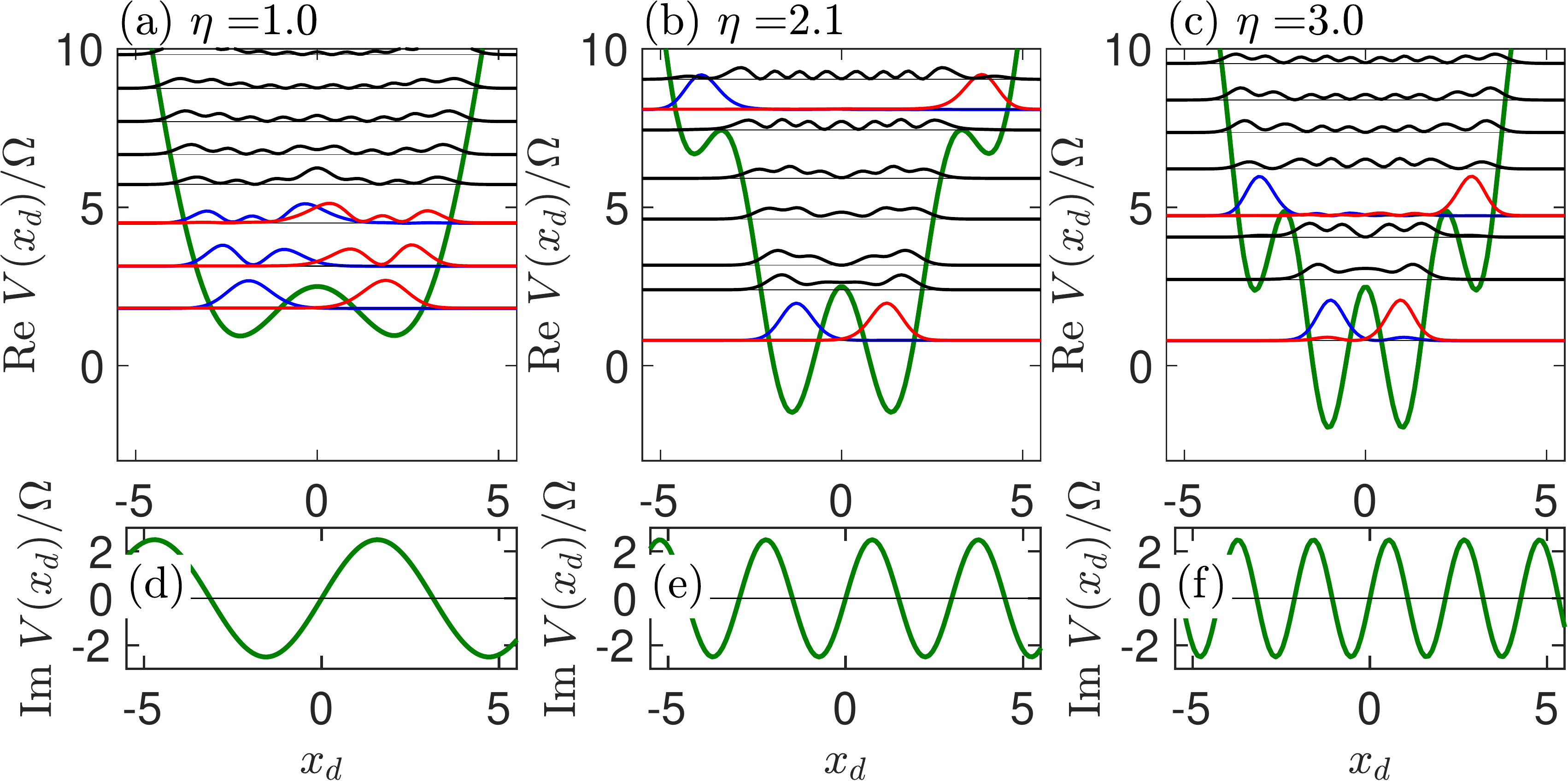}
    \caption{(a--c) 
    Eigenfunctions $|\psi^2_n(x_d)|$ of the full Hamiltonian Eq.~\eqref{eq:H_d}  calculated taking into account also the dissipative coupling  (black, red and blue curves), the potential $V(x_d)$ (green curve) and the energy levels (thin black horizontal lines), corresponding to three specific values of optomechanical coupling strength $\eta$. The states with large (small) negative values of the imaginary energy, which are localized at positive (negative) $x_d$ are shown by blue (red) color.  (d--e) Imaginary parts of the potential.
    Calculation parameters are the same as in Fig.~\ref{fig:4}.}
    \label{fig:5}
\end{figure}

We now proceed to include the dissipative optomechanical coupling. Figure~\ref{fig:4}, similarly to Fig.~\ref{fig:2}, shows the dependence of the energy levels on the optomechanical coupling strength. It is calculated by numerical diagonalization of the full Hamiltonian Eq.~\eqref{eq:H_d}, taking into account also the imaginary part of the potential. Similarly to Fig.~\ref{fig:2}, we also show the quasiclassical Bohr-Sommerfeld quantization phase $\{n \}$ and the   approximate equations Eq.~\eqref{eq:bohr1} for the energies. The real parts of the energies, shown in Fig.~\ref{fig:4}(a), behave in approximately the same way as in Fig.~\ref{fig:2}: optomechanical coupling induces periodic modulation and also leads to appearance  of additional levels if the coupling strength $\eta$ is above the threshold. However, in contrast to the Hermitian case, the eigenenergies can now acquire an imaginary contribution, which is shown in  in Fig.~\ref{fig:4}(b).

The potential $V(x)$ in Eq.~\eqref{eq:H_d} is parity-time ($\mathcal PT$) symmetric~\cite{Iorsh2020}. If the $\mathcal PT$ symmetry persists for the wavefunctions, the eigenenergies are real. However, the $\mathcal PT$ symmetry for the wavefunctions can  spontaneously break in a certain range of parameters. In that case, a pair of states with complex-conjugated eigenenergies  and the wavefunctions related by the $\mathcal PT$ operation appear in the spectrum. 
 This $\mathcal PT$ symmetry breaking is most evident for the localized states shown by red and blue curves  in Fig.~\ref{fig:5}(b,c). For  the Hermitian problem [Fig.~\ref{fig:3}(b,c)], each state of the localized states pair posses $\mathcal P$ (mirror) symmetry. The imaginary part of the potential $V''(x_d) = \rmi \sin \eta x_d$ breaks the $\mathcal P$ symmetry. The energies of the states pair are split by an imaginary value, and the wavefunctions become localized either at positive or negative values of the coordinate $x_d$.

\section{Summary}\label{sec:summary}

To summarize, we have developed a theory for interaction of two vibrating atoms, trapped near the waveguide, with the photons, propagating  inside the waveguide. We have started by obtaining an effective Hamiltonian for the coupled vibrations of the atoms, in the regime when they are excited by a single polariton. Next, we have diagonalized this Hamiltonian numerically (i) neglecting and (ii) including dissipative optomechanical coupling. We have focused on the semiclassical regime with a relatively large number of vibration quanta $n\sim 10$. In both cases we find spatially localized modes of the atom vibrations. This localization is driven by a strong optomechanical interaction and can be interpreted semiclassically, by using  the Bohr-Sommerfeld quantization rule for the vibrations. When the dissipative optomechanical coupling is taken into account, the pairs of mirror-symmetric localized modes exhibit $\mathcal PT$-symmetry breaking, and become localized either at positive or negative relative atom motion coordinate.

Our results demonstrate the potential of  hybridized quantum systems, based either on atoms or superconducting qubits, where excitations of different nature are coupled to each other,  to control quantum light-matter correlations.  Even richer   physics might be unraveled  in the regime with many atoms or many atomic excitations that will be hopefully accessible in the near future experiments.

\acknowledgements
The authors acknowledge useful discussion with M.V.~Perel.
This work has been supported by the  Russian Science Foundation Grant No. 20-12-00194.

%

\end{document}